\magnification=\magstep1
\voffset =-2 truecm
\hoffset =0 truecm
\vsize = 9.5 truein
\hsize = 6.4 truein
\baselineskip =18 pt
\vskip 20 pt
\centerline {\bf BARYON DECUPLET MASSES FROM THE VIEWPOINT OF $q$-EQUIDISTANCE}

\vskip 15 pt
\centerline {\bf A.M.Gavrilik,\ I.I.Kachurik,\  A.V.Tertychnyj}
\vskip 5 pt
\centerline {Institute for Theoretical Physics, Academy of Sciences of Ukraine}
\centerline {(03143 Kiev-143, Metrologichna str., 14b)}
\vskip 10 pt
\noindent {\sl Summary }

\noindent {Masses of baryons ${3\over 2}^+$
are calculated  on the base of representations of dynamical
"pseudounitary" $q$-deformed algebra $u(4,1)_q$ which provides
necessary breaking of the 4-flavor symmetry realized by the assumed
$q$-algebra $su(4)_q$. It is demonstrated that, contrary to the case
of $su(3)_q$-octet baryons ${1\over 2}^+$, one and the same
$q$-analog of mass relation for baryons ${3\over 2}^+$ from decuplet
embedded into ${\it 20}$-plet of $su(4)_q$ follows from evaluations
within all the different admissible "dynamical" representations.

\vskip 15 pt

{\bf 1.} Quantum algebras (the $q$-algebra $su(2)_q$ as most popular
among them) introduced in the eighties in the context of integrable
systems, now occupy a very important place both in pure mathematical
disciplines (e.g., such as theory of knots and links) and in
mathematical as well as theoretical physics. A couple of years
ago $su(2)_q$ has appeared in the framework of some interesting
phenomenological models aimed to describe rotational spectra
of (super)deformed heavy nuclei and diatomic molecules [1].
Concerning miscellaneous applications of quantum groups/$q$-algebras
in a more wide context let us quote refs. [2-4] and references
given therein.

In attempts to find phenomenological application of higher rank
quantum algebras, one may encounter some new features absent in
$su(2)_q$ case. Among such features we mention the necessity to
deal with nonsimple-root elements (hence, with $q$-Serre relations)
of those algebras, nontriviality of the concepts of (and formulas
for) Casimir operators, Clebsch-Gordan coefficients (CGC's), etc.

    Recently, the use of higher rank quantum algebras $su(n)_q$
(or $u(n)_q$) in order to replace conventional unitary groups
$SU(n)$ and their irreducible representations (irreps) in
describing flavor symmetries of hadrons (vector mesons $1^-$ and
baryons ${1\over 2}^+$) has been proposed [5-7]. With the help of
the corresponding algebras $u(n+1)_q$ or $u(n,1)_q$
of "dynamical" symmetry, one can realize necessary breaking of
$n$-flavor symmetries up to exact (for strong interactions alone)
isospin symmetry $su_q(2)_I$ and obtain some $q$-analogs
of mass relations (MR's). Such an application of the $q$-algebras
$u(n)_q$ for obtaining $q$-analogs of hadron MR's
uses generators that correspond to both simple-root
elements and nonsimple-root ones (thus, the $q$-Serre relations
play definite role [6]); from another side, it exploits rather
simple model which allows one to circumvent difficulties
related with $q$-CGC's and $q$-Casimirs.

As implied by the results of [5-7], replacement of the conventional
unitary groups of hadronic flavor symmetries by their quantum
counterparts may also lead to further interesting consequences.
Let us sketch some of them briefly.

    In the case of the ordinary (non-deformed) $SU(n)$ symmetry
of hadrons with $n$ quark flavors, the approach based on dynamical
unitary groups allows one to obtain [8] besides the well-known
octet mass sum rule (MSR) $ 3m_{{\omega}_8} + m_{\rho} = 4 m_{K^*} $,
also the higher-flavor set of MSR's for vector mesons
$1^-$. A comparison of the octet MSR with existing data is
impossible without introducing certain mixing between isosinglet
${\omega }_8$ and $SU(3)$ singlet (i.e. ${\omega }_8$ must be
considered as a superposition of realistic mesons $\phi $ and $\omega $)
with some mixing angle to be determined from empirical data.
Likewise, in cases of more flavors ($n>3$) one needs $n-2$
mixing angles. But, extending the approach of [8] to $q$-algebras $su(n)_q$,
one can derive for vector mesons the q-deformed
MR's which admit (at $\vert q\vert =1,\ q\ne \pm 1$) some principally
different (from the $q=1$ case) treatment which {\it allows to avoid}
[5,6] manifest introducing of singlet mixing}.

Next, it turns out that all the $q$-dependence in vector meson
masses and in coefficients of their MR's is expressible in terms
of certain Lorant-type polynomials (of variable $q+q^{-1}$) which
were noticed to be related [6] with some knot invariants (Alexander
polynomials $P_{n}(q)$~). In a sense, the polynomial $P_{n}(q)$
through its root $q(n)$ enables one to determine the strength of
deformation at every fixed $n$, and due this property may be
called a {\it defining polynomial} for the corresponding vector
meson MSR. This way is principally different from the choice of $q$
by fitting procedure [1,4].
Further, by utilizing the {\it quantum} groups/$q$-algebras instead of
conventional unitary groups of flavor symmetries,
together with 'dynamical' {\it quantum} algebras,
we get as a result that the collection of torus knots is put into
correspondence [6] with heavy vector quarkonia. In other words,
application of the embeddings $u(n)_q\subset u(n+1)_q$ to analysis
of (vector) meson masses opens an appealing possibility of definite
{\it topological characterization of heavy flavors}, since
the number $n$ just corresponds to  $2n\!-\!1$ overcrossings of
2-strand braids whose closures give those $(2n-1)$-torus knots.

    The approach of [5,6] was recently extended in order to treat
the case of baryons ${1\over 2}^+$ (including charmed ones), by
adopting again the algebra $u(4)_q$ for the 4-flavor symmetry [7].
Unlike the case of vector mesons where 'compact' $q$-algebras
$u(n+1)_q$ were used for dynamical symmetry, and in some analogy
to the case of baryon MR's obtained with non-deformed
dynamical $u(4,1)$-symmetry [8], it was more convenient there
to exploit representations of the 'noncompact' dynamical
symmetry, realized by the $q$-algebra $u(4,1)_q$ in order
to effect necessary symmetry breakings.

    On the base of evaluations within certain concrete irrep
it was demonstrated [7] that the resulting $q$-analog of
baryon octet MR yields either the usual
Gell-Mann--Okubo (GMO) mass sum rule [9]
$m_N+m_{\Xi}={3\over 2}m_{\Lambda}+{1\over 2}m_{\Sigma } $
or a very successful novel MSR if one fixes
respectively $q=1$\ or $q=e^{i\pi \over 6}$. These values
are the roots of one and the same {\it defining polynomial}
$A_q$ appearing in the $q$-analog.
Another $q$-analogs (with different defining $q$-polynomials)
of octet MR can be obtained by performing calculations within another
specific representations of $u(4,1)_q$. Therefore, one can say that
the use of quantum algebras in place of non-deformed ones, in a sense,
"removes degeneracy".

The purpose of the present letter is to apply the scheme
developped in [7] to the case of baryons ${3\over 2}^+$ which
constitute the decuplet irrep of $su(3)_q$. According to
conventional descriptions, symmetry breaking results in equidistant
differences between masses of isoplet members of the decuplet [9].
However, the empirical data show that actually there is a deviation
from exact equidistance. Use of the $q$-algebras $su(n)_q$ instead of
ordinary $SU(n)$ seems to provide natural and rather simple accounting
of such a deviation. After presenting the results of evaluation
of decuplet baryon masses in two specific irreps of the dynamical
$u(4,1)_q$, an assertion is proved which states that certain
$q$-analog of MR, see eq. (5) below, holds in all the admissible
dynamical representations. Also, it is argued that the deformation
parameter  $q$  must be pure phase.
\smallskip

{\bf 2.} As already mentioned, among possible approaches to
$SU(n)_q$-symmetry breaking necessary in order to obtain mass
relations for hadrons of $n$ quark flavors, we prefer the scheme
which was used in ref.[7] for the case of baryons ${1\over 2}^+$ and
which is nothing but a straightforward extension to quantum groups
of the approach based on noncompact (pseudo-unitary) dynamical
groups, see [8] and references given therein.

In what follows, we use the 20-dimensional irrep $T_{20}$
of $u(4)_q$ (instead of $su(4)_q$ ) which is characterized by
the 4-tuple $\{p+2,p-2,p-3,p-4\}$ with some $p\in {\bf Z}$
(explicit value of $p$ will not enter final expressions).

To form state vectors for baryons ${3\over 2}^+$ from
decuplet, we use the (orthonornalized) Gel'fand-Zetlin basis
elements constructed in accordance with the chain
$u(4,1)_q\supset u(4)_q\supset u(3)_q\supset u(2)_q$,
namely,
$$ \vert B_i\rangle\leftarrow\!\rightarrow\vert \chi; {\bf m}_4,
{\bf m}_3, {\bf m}_{2(i)}, Q_i\rangle ,\ \ i=1,...,4, \eqno (1)
$$
 where $\chi\equiv\{l_1,l_2,l_3;c_1,c_2\}$ labels irreps of
$u(4,1)_q$; ${\bf m}_4\equiv\{p+2,p-2,p-3,p-4\}$ and
${\bf m}_3\equiv\{p+2,p-2,p-3\}$ label respectively ${\it 20}$-plet
of $u(4)_q$ and ${\bf 10}$-plet of its subalgebra $u(3)_q$;
${\bf m}_{2(i)},$\ with $i=1,..,4$,\ characterizes
isoplets (of dimensions $1,...,4,$)\ from the ${\bf 10}$-plet
(see ref.[7] for more details concerning the approach and irreps
of $u(4,1)_q$~).

Another ingredient of this scheme, the mass operator, is
composed of "noncompact" generators of the $q$-algebra $u(4,1)_q$
and is taken in the form [7]
$$\hat {M_4} = M_o + a I_{45}I_{54} + b I_{54}I_{45}\hskip 120 pt $$
 $$ + \alpha I_{35}\tilde I_{53} +
  \beta\tilde I_{53} I_{35} +\tilde \alpha\tilde I_{35} I_{53} +
  \tilde \beta I_{53}\tilde I_{35}.\ \  \eqno (2)$$
Since $I_{5i},\ I_{i5}$\ with $i=1,2$ and their dual (tilded) counterparts
are absent in eq.(2), this operator  commutes with "isospin" $su_q(2)_I$.
To reduce the number of independent parameters we impose
${\alpha}_i =\tilde {\alpha}_i , \ \ {\beta}_i =\tilde {\beta}_i$.

The representation $T_{\chi}$ of 'dynamical' algebra is called
$T_{20}$-{\it compatible} if it contains irrep $T_{20}$
of its subalgebra $u(4)_q$. From definitions of integer-type
infinitesimally unitary irreps of $u(4,1)_q$ which
include the corresponding branching rules under restriction
$u_q(4,1)\vert_{u_q(4)}$, it is an easy task to check that
the following assertion is valid.

\smallskip
{\bf Proposition 1}. The integer type irreps of $u(4,1)_q$ which are
$T_{20}$-compatible are contained in the following list
(here $s\in {\bf Z}_+$, and either $t=0$ or $t=1$):
\medskip
\noindent $(i)\ D_-^{12}(p-1,p-3,p-4;p+1+s,p-2)$\ and
\ $D_-^{12}(p+t,p-3,p-4;p+1+s,p-1+t)$;

\noindent $(ii)\ \tilde D_+^{24}(p,p-3,p-4;p-3,p-5)$\ and
\ $\tilde D_+^{24}(p-1,p-3,p-4;p-3,p-5)$;

\noindent $(iii)\ \tilde D_+^{34}(p-1,p-3,p-4;p-4,p-5)$\ and
\ $\tilde D_+^{34}(p,p-3,p-4;p-4,p-5)$;

\noindent $(iv)\ \tilde D_+^{11}(p-1,p-3,p-4;p-1,p)$;

\noindent $(v)\ D_+^4(p,p-3,p-4;p-4-s,p-4-s)$\ and
\ $D_+^4(p-1,p-3,p-4;p-4-s,p-4-s)$;

\noindent $(vi)\ D_-^1(p,p-3,p-4;p+1+s,p+1+s)$\ and
\ $D_-^1(p-1,p-3,p-4;p+1+s,p+1+s)$.

\medskip
{\bf 3}. We are now in a position to obtain the ($q$-dependent)
expressions for masses of the hadrons under consideration. Using first
the irrep $D_-^{12}(p-1,p-3,p-4;p+2,p-2)$ we calculate
$\langle B_i\vert \hat M_4\vert B_i\rangle$,\ $i=1,...,4$\ , and
obtain the following results for baryon ${3\over 2}^+$  (iso-quartet,
-triplet, -doublet, and -singlet) masses from {\bf 10}-plet of $su(3)_q$
(here $[k]\equiv[k]_q\equiv{q^k-q^{-k}\over q-q^{-1}}$):
$$\eqalignno {m_{{\Delta}_4}&=m_{\bf 10}+a[4]+\alpha [2][4], &{}\cr
              m_{{\Sigma}_3}&=m_{\bf 10}+a[4]+\alpha\{[2]^2[4]
+{[2]\over [6]}([2]+[4][5])-4[4]\}, &{}\cr
              m_{{\Xi}_2}&=m_{\bf 10}+a[4]+\alpha\Bigl \{[2][3][4]
+{[2]^2\over [6]}([2]+[4][5])-4[2][4]\Bigr \}, &{(3)}\cr
              m_{{\Omega}_1}&=m_{\bf 10}+a[4]+\alpha\Bigl \{[2][4]^2
+{[2][3]\over [6]}([2]+[4][5])-4[3][4]\Bigr \}. &{} }$$
We observe that the parameters $b$ and $\beta$ (see (2)~) do not enter
the above expressions, and this fact is a characteristic feature
of the representation just exploited. Indeed, in this specific
irrep there is no freedom to raise (or to lower) the $l$-coordinates
$l_{14}$,\ $l_{24}$,\ $l_{34}$ in the highest weight ${\bf m}_4$ of
its $u(4)_q$-subrepresentation since that is forbidden by
$u_q(4,1)\vert_{u_q(4)}$-branching rules (the corresponding matrix
elements of representation operators $I_{45}, I_{54}$ do vanish).
Only $l_{44}$ can be lowered.

In terms of the denotion
$$m=m_{\bf 10}+a[4],\quad\quad {\alpha}'=\alpha [2][4],
\quad\quad {\alpha}''=\alpha\Bigl\{{[2]\over [6]}([2]+[4][5])-4[4]\Bigr\}$$
the expressions (3) take essentially more transparent form, namely
$$\eqalignno {m_{{\Delta}_4}&=m+{\alpha}', &{}\cr
              m_{{\Sigma}_3}&=m+{\alpha}'[2]+{\alpha}'', &{}\cr
              m_{{\Xi}_2}&=m+{\alpha}'[3]+{\alpha}''[2], &{(3')}\cr
              m_{{\Omega}_1}&=m+{\alpha}'[4]+{\alpha}''[3].&{} }$$
The $q$-equidistance (it generalizes usual equidistance of the case of
$SU(n)$-symmetries) is manifest here. In the 'classical' (non-deformed)
limit $q\to 1$, the conventional equidistance (with the 'step' or
distance equal to $({\alpha}'+{\alpha}''){\vert}_{q=1}=-{2\over 3}\alpha$),
is recovered:
$$m_{{\Omega}_1}-m_{{\Xi}_2}=m_{{\Xi}_2}-m_{{\Sigma}_3}
=m_{{\Sigma}_3}-m_{{\Delta}_4}= -{2\over 3}\alpha \eqno (4)$$
(recall that it was the equidistance relation (4) that led to
prediction and discovery of the famous  $\Omega^-$-particle [9]).
\smallskip
{\bf 4.} To make some contact with empirical situation in the
$q$-deformed case, it is important to fix the parameter $q$
appropriately. Unfortunately, at the moment we have not a conclusive
idea of how to do this (contrary to the cases of vector mesons [5-6]
and octet baryons ${1\over 2}^+$, see [7]). Nevertheless, we can
decide which of the two options,

$(a)\ \ \ q\in {\bf R}$\qquad \qquad or \qquad \quad $(b)\ \ \ q=e^{ih}, \ \ \ h\in {\bf R}$,

\noindent is realized. To this end, we deduce from (3) or (3') the
relation for masses which does not
contain undetermined parameters. From the differences
$m_{{\Sigma}_3}-m_{{\Delta}_4},$\ \ $m_{{\Xi}_2}-m_{{\Sigma}_3},$\ \ and
$m_{{\Omega}_1}-m_{{\Xi}_2}$  we form a combination that results in
the desired mass relation (it is of the "$q$-average" type):
$$
{m_{{\Sigma}_3}-m_{{\Delta}_4}+m_{{\Omega}_1}-m_{{\Xi}_2}\over [2]_q}=
m_{{\Xi}_2}-m_{{\Sigma}_3}. \eqno (5)  $$
The dependence on deformation parameter here is the simplest possible
one. Now recall that empirical values for the participating
baryon masses averaged over isoplets are [10]
$m_{{\Delta}_4}=1232$ Mev, \ $m_{{\Sigma}_3}=1384.6$ Mev, \ $m_{{\Xi}_2}=
1533.4$ Mev,  and  $m_{{\Omega}_1}=1672.4$ Mev. Substitution into eq.(5) of
the empirical data shows that we must have $[2]_q\approx 1.95$. Since at
real $q$ other than 1 necessarily $[2]_q\equiv q+q^{-1}>2,$  we conclude
definitively in favor of the option $(b)$.

\smallskip
{\bf 5.} Now let us go over to exploiting another 'dynamical'
representation, namely, $D_-^{12}(p,p-3,p-4;p+2,p-1)$. Performing
necessary evaluations within it, this time we obtain the following
expressions for isoplet masses:
$$\eqalignno {m_{\Delta_4}&=
m +\alpha {[2][4][5]\over [3]}
       +\beta {[2]^2\over [3]}, &{}\cr
               m_{\Sigma_3}&=
m +\alpha\Bigl \{ {[2]\over [3]}\Bigl ( {[2]+[4][5]^2\over [6]}+[4]^2
+[4][6]\Bigr ) -4{[4][5]\over [3]}\Bigr \} +\beta {[2]^3\over [3]^2}, &{}\cr
                 m_{{\Xi}_2}&=
m +\alpha\Bigl \{ {[2]\over [3]}\Bigl ( {[2]^2+[2][4][5]^2\over [6]}
+[2][4][6]\Bigr ) +[2][4]-4{[2][4][5]\over [3]}\Bigr \} +
\beta {[2]^2\over [3]^2},                          &{(6)}\cr
              m_{{\Omega}_1}&=
m +\alpha\Bigl \{ [2]\Bigl ( {[2]+[4][5]^2\over [6]}
+[4][6]+{[2][4]\over [3]}\Bigr ) -4[4][5]\Bigr \}   &{} }$$
where $m =m_{\bf 10}+{[4][5]\over [3]}a+{[2]\over [3]}b$.
Classically, again we are led to the equidistance (the 'step' now
equals to  $-2\alpha -{4\over 9}\beta $).

At $q\ne 1$ the expressions (6) apparently differ from
eqs.(3). Indeed, now masses depend both on the parameter $\alpha$\
and on the parameter $\beta$. Furthermore, introducing the denotion
$A=\alpha {[2][4]\over [3]}$,\ \
$A'=\alpha\bigl\{ [2]\bigl ( {[2]+[4][5]^2\over [3][6]}
+{[4][6]\over [3]}\bigr ) -4{[4][5]\over [3]}\bigr\} $,\ \ and
$B=\beta {[2]^2\over [3]^2}$,\ we rewright eq.(6) in the form
$$\eqalignno {m_{\Delta_4}&=m + [5]A + [3]B, &{}\cr
              m_{\Sigma_3}&=m + [4]A + A'+ [2]B, &{}\cr
                 m_{\Xi_2}&=m + [3]A + [2]A' + B, &{(6')}\cr
              m_{\Omega_1}&=m + [2]A + [3]A',&{} }$$
and observe here both decreasing and growing sequences of $q$-numbers
simultaneously in the coefficients when reading from the top down
({\it generalized} $q$-{\it equidistance}).

Like before, we are interested in obtaining the mass relation
which is independent of undetermined parameters $m,\ \alpha$\ and
\ $\beta$. Forming the differences $m_{{\Sigma}_3}-m_{{\Delta}_4},$\ \
$m_{{\Xi}_2}-m_{{\Sigma}_3},$\ \ and  $m_{{\Omega}_1}-m_{{\Xi}_2}$
we arrive at the relation which {\it coincides} with eq.(5).

Is such a coincidence merely an accidental fact, or may be the
$q$-relation (5) is in a sense universal (i.e. holds independently
of the choice of dynamical representation)? The following assertion
gives the answer.

\smallskip
{\bf Proposition 2.} The $q$-average mass relation (5) for isoplet
masses from the {\bf 10}-plet of $su(3)_q$ is valid in any of the
$T_{20}$-compatible irreps $T_{\chi}$ from the list given
above (see proposition 1).

To prove this, we have to calculate the matrix elements
$\langle i\vert \hat M_4\vert i\rangle$ for every isoplet $B_i$
( $i$ runs over ${\Delta}_4$,\ \ ${\Sigma}_3$,\ \ ${\Xi}_2$,
\  and \ ${\Omega}_1$) within the ${\bf 10}$-plet of $su(3)_q$
embedded into (an arbitrary) $T_{20}$-compatible irrep $T_{\chi}$.

From the explicit action formulas for the representation operators
$T_{\chi}(I_{n,n+1})$ and $T_{\chi}(I_{n+1,n})$ we infer two facts.
First, in arbitrary representation $T_{\chi}$, matrix elements
of the invariant operator $\hat M_0$ and of the terms $I_{54}I_{45}$,
$I_{45}I_{54}$ (see eq. (2) ) are the same for all the isoplets $B_i$.
Denote these as $\langle i\vert (\hat M_0 + a I_{45}I_{54}
+ b I_{54}I_{45}) \vert i\rangle \equiv F_{\chi }$.
Second, the dependence of $\langle i\vert \hat M_4\vert i\rangle $
on the signature of specific isoplet $B_i$ has its origin only in
the action of the operators $I_{34}$ and $I_{43}$ which
compose the operators $I_{35}, \tilde I_{35}, I_{53}, \tilde I_{53}$.
Moreover, denoting ${\langle A_{53}\rangle}_i\equiv
\langle i\vert (I_{53}\tilde I_{35}+\tilde I_{53}I_{35})\vert i\rangle$
and ${\langle A_{35}\rangle}_i\equiv
\langle i\vert (I_{35}\tilde I_{53}+\tilde I_{35}I_{53})\vert i\rangle$,
it is not difficult to show that
$$\eqalign { {\langle A_{53}\rangle}_i &= A_1\vert [l_{13}-p+2-i]\vert +
A_2 \vert [l_{33}-p+2-i]\vert =  A_1[k]+A_2[u],\cr
{\langle A_{35}\rangle}_i &= C_1\vert [l_{13}+1-p+2-i]\vert +
C_2 \vert [l_{23}+1-p+2-i]\vert = C_1[k+1]+C_2[u-2], }$$
where $k=4-i,\ u=1+i,\ (i=1,...,4)$, and the coefficients $A_1,\ A_2,
\ C_1,$\ and $C_2$  depend on ${\chi }$, on the signature ${\bf m}_4$ of $20$-plet,
on the signature ${\bf m}_3$ of the ${\bf 10}$-plet, but do not
depend on particular isoplet $B_i$.

From the calculated expressions for isoplet masses
we find the differences:
$$\eqalign { m_{\Sigma_3}-m_{\Delta_4}&=A_1([k+1]-[k])+A_2([u-1]-[u])\cr
&+C_1([k+2]-[k+1])+C_2([u-3]-[u-2]),\cr
m_{\Xi_2}-m_{\Sigma_3}&=A_1([k+2]-[k+1])+A_2([u-2]-[u-1])\cr
&+C_1([k+3]-[k+2])+C_2([u-4]-[u-3]),\cr
m_{\Omega_1}-m_{\Xi_2}&=A_1([k+3]-[k+2])+A_2([u-3]-[u-2])\cr
&+C_1([k+4]-[k+3])+C_2([u-5]-[u-4]). }$$
Finally, with the  $q$-number identity  $[n+2]=[2][n+1]-[n]$\ taken
into account, the relation (5) follows and this completes the proof.

\smallskip
{\bf 6}. Summarizing, we have shown that, unlike the case of
$SU_q(3)$-octet baryons ${1\over 2}^+$ (see [7]), in the present
case of decuplet baryons ${3\over 2}^+$  the obtained $q$-average
mass relation (5) holds for any of $T_{20}$-compatible
irreps $T_{\chi}$ of "dynamical" $q$-algebra $u(4,1)_q$.

Taking into account the empirical values of baryon masses we have
made a conclusion that the deformation parameter $q$ must be a pure
phase (see the option $(b)$ in sec.4) in order that the relation (5)
would give realistic mass sum rule. However, the problem of
more exact fixation of the deformation parameter remains open
since it depends drastically on an improvement of the accuracy of
data for empirical masses (remark that it is the isoquartet masses
the values of which are most ambiguous).
\medskip
This research was partially supported by the International Science
Foundation under the grant U4J000 and by the Ukrainian State
Foundation for Fundamental Research.
\bigskip
\noindent \ \ {\bf References}
\medskip
\item {[1]}Iwao S 1990 {\sl Progr. Theor. Phys.} {\bf 83} 363.
\item {}Raychev P P, Roussev R P and Smirnov Yu F 1990
      {\sl J. Phys. G} {\bf 16} 137.
\item {}Bonatsos D et al. 1990 {\sl Phys. Lett.} {\bf 251B} 477.
\item {[2]}Biedenharn L C 1990 {\sl An overview of quantum groups}, in:
      Proc. $XVII^{th}$ Int. Coll. on Group Theor. Methods in Physics
      (Moscow, June 1990).
\item {[3]}Zachos C 1991 {\sl Paradigms of quantum algebras}, in:
      Symmetry in Science V (B Gruber, L C Biedenharn, and
      H D Doebner, eds.), pp 593-609.
\item {[4]}Kibler M 1993 {\sl Lyon preprint} LYCEN/9358.
\item {[5]}Gavrilik A M 1993 {\sl Physics in Ukraine. Quantum Fields
      and Elementary Particles} (Proc. Int. Conf., Kiev);
      Gavrilik A M, Tertychnyj A V {\sl Kiev preprint} ITP-93-19E.
\item {[6]}Gavrilik A M 1994 {\sl J. Phys. A: Math. and Gen.} {\bf 27} L91.
\item {[7]}Gavrilik A M, Kachurik I I, Tertychnyj A V {\sl Kiev preprint}
           ITP-94--34E
\item {}[\tt hep-ph/9504233].
\item {[8]}Gavrilik A M, Shirokov V A 1978 {\sl Yadernaya Fizika} {\bf 28} 199.
\item {}Gavrilik A M, Klimyk A U 1989 {\sl Symposia Mathematica} {\bf 31} 127.
\item {[9]}Novozhilov Yu V 1975 {\sl Introduction to elementary
       particle theory} (New York:
\item {} Pergamon).
\item {}Gasiorowicz S 1966 {\sl Elementary particle theory}
      (New York: J Willey).
\item {[10]}Particle Data Group 1990 {\sl Phys. Lett. B} {\bf 239} 1.

\qquad\qquad\qquad\qquad\qquad\qquad\quad
\qquad\qquad\qquad\qquad\qquad\qquad                   { Received 01.03.95}

\end